\documentclass[epj]{webofc}
\usepackage[varg]{txfonts}   % Web of Conferences font
\usepackage{wrapfig}
\usepackage{gensymb}
\usepackage{lineno}
\wocname{EPJ Web of Conferences}
\woctitle{New Frontiers in Physics 2014}
\begin{document}
\selectlanguage{english}
\title{Recent results from Daya Bay experiment}
%
% subtitle is optional

\author{Dmitry V.Naumov (on behalf of the Daya Bay collaboration)\inst{1}\fnsep\thanks{\email{dnaumov@jinr.ru}}}

\institute{Joliot-Curie, 6, Dubna, Russia, 141980}

\abstract{%
  This manuscript is a short summary of my talk given at ICNFP2014 Conference. Here we report on new results of 
$\sin^22\theta_{13}$ and $\Delta m^2_\text{ee}$ measurements, search for the sterile neutrino within $10^{-3} 
\text{ eV}^2 <\Delta m^2_{41}<0.1\text{ eV}^2$ domain and precise measurement of the reactor absolute antineutrino flux.
}
\maketitle
\section{Introduction}
\label{intro}
Generations of leptons (and similarly quarks) mix in their interactions with $W^\pm$ bosons within the Standard Model. 
The mixing is governed by a unitary matrix of dimension equal to the numer of lepton families. For three 
lepton's generations assuming the Dirac nature of neutrino the corresponding $3\times 3$ 
Pontecorvo-Maki-Nakagawa-Sakata (PMNS) matrix is conviniently described by three mixing angles $\theta_{12}$, 
$\theta_{23}$ and $\theta_{13}$ and one CP-violating phase $\delta$. Two more CP-violating phases are required to 
describe the Majorana neutrinos. 

The mixing in the lepton sector leads to a spectacular phenomenon of flavour transformation over time and space 
-- the so called neutrino oscillations first described by Pontecorvo~\cite{Pontecorvo:1957cp,Pontecorvo:1967fh}.  
Nowdays the theory of neutrino oscillations is rather well established in both relativistic quantum 
mechanics and within the framework of quantum field 
theory~\cite{Beuthe:2002ej,Cardall:1999ze,Stockinger:2000sk,Grimus:1999ra,Grimus:1999zp,Grimus:1999pm,Akhmedov:2007fk, 
Akhmedov:2012mk,Akhmedov:2010ms,Naumov:2010um,Naumov:2009zza}.

Experimentally neutrino oscillations is also a well established phenomenon. Two mixing angles $\theta_{12}$ and 
$\theta_{23}$ are accurately measured by solar, reactor, atmospheric and accelerator neutrino 
experiments~\cite{Abe:2008aa,Adamson:2013ue,Ashie:2004mr} assuming neutrino oscillation hypothesis as an explanation of 
observed rates of appearance and disappearance of neutrino flavours. Under the same hypothesis two mass squared 
differences are measured as well. First is $\Delta m^2_{21}$ and second is  $\Delta m^2_{\mu\mu}$ which is a flavour 
mixture of $\Delta m^2_{31}$ and $\Delta m^2_{32}$ since current experiments have little sensitivity to the mass 
hierarchy. Atmospheric and accelerator neutrino experiments measure   $\Delta m^2_{\mu\mu} \simeq 
\sin^2\theta_{12}\Delta m^2_{31} + \cos^2\theta_{12}\Delta m^2_{32} + 
2\Delta m^2_{21}\sin\theta_{12}\cos\theta_{12}\sin\theta_{13}\tan{\theta_{23}}\cos\delta$~\cite{Nunokawa:2005nx}. Till 
2012 the value of $\theta_{13}$ was unknown. A general feeling was that this angle could be very small as Chooz 
experiment provided an upper limit $\sin^22\theta_{13}<0.15$~\cite{Apollonio:2002gd}. This angle could be measured by 
both accelerator and reactor neutrinos and a number of experiments (MINOS, T2K, Double Chooz, RENO and Daya Bay) began a 
race for it. First indications for non zero value of $\theta_{13}$ came from MINOS~\cite{Adamson:2011qu}, 
T2K~\cite{Abe:2011sj} and Double Chooz~\cite{Abe:2011fz} in 2011. Also the global analysis of solar and KamLAND data 
indicated for non-zero value of 
$\theta_{13}$~\cite{Fogli2011}. However none of these indications reached a significance even of three standard 
deviations. 

The discovery of non zero value of $\theta_{13}$ was done by a reactor experiment Daya Bay which observed a deficit of 
$\bar{\nu}_e$ flux at the far site $R=0.940\pm 0.011 \text{ (stat) } \pm 0.004 \text{ (syst) }$ which can be explained 
due to neutrino oscillations with $\sin^22\theta_{13}=0.092\pm 0.016 \text{ (stat) } \pm 0.005 \text{ (syst) 
}$~\cite{An:2012eh} in a three-neutrino framework. Soon after RENO Collaboration confirmed this 
result~\cite{Ahn:2012nd}. A solid determination of a relatively large value of $\theta_{13}\simeq 9^{\degree}$ opened 
possibilities to study the neutrino mass hierarchy and $\delta$. 

In what follows the most recent results of the Daya Bay Collaboration are reviewed. For the yet unpublished results 
please refer~\cite{Chao, Weili}. In Sec.~\ref{sec:dayabay_experiment} we briefly review the Daya Bay detector and 
energy model. Please refer the following papers~\cite{An:2013zwz,An:2014ehw} for the event selection. In 
Sec.~\ref{sec:results} we report on new Daya Bay results of neutrino oscillations, sterile neutrino search, measurement 
of reactor antineutrino flux. Finally, in Sec.~\ref{sec:summary} we draw our conclusions.

\section{Day Bay experiment}
\label{sec:dayabay_experiment}
\begin{figure}[htb]
 \begin{tabular}{cc}
 \centering\includegraphics[width=0.45\textwidth,clip]{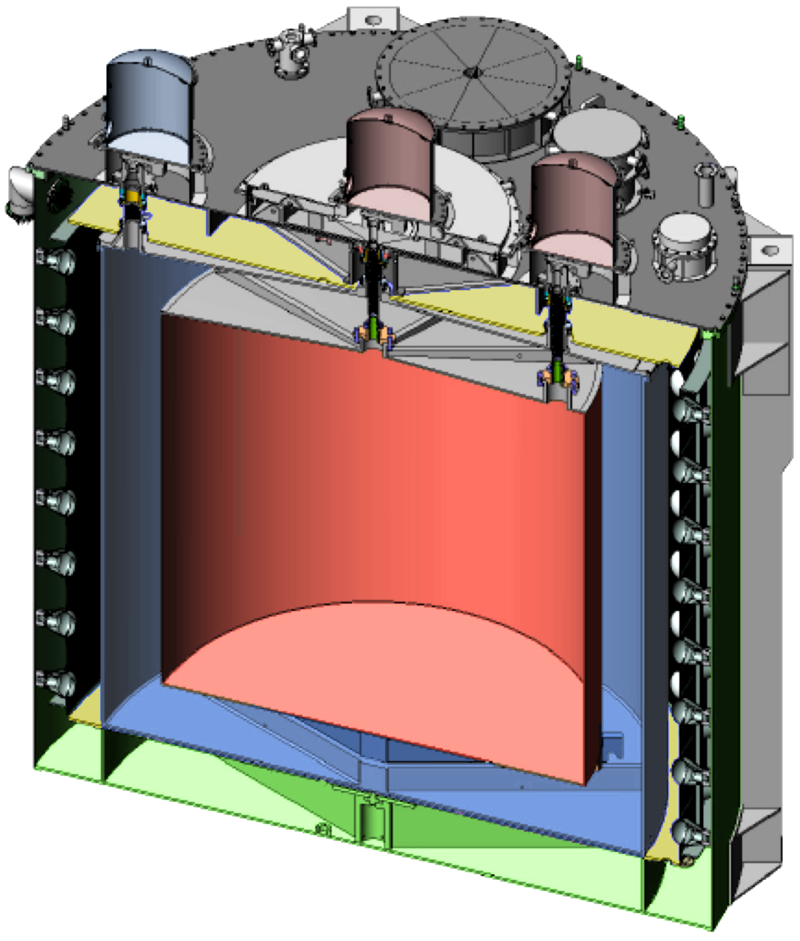}&
 \centering\includegraphics[width=0.45\textwidth]{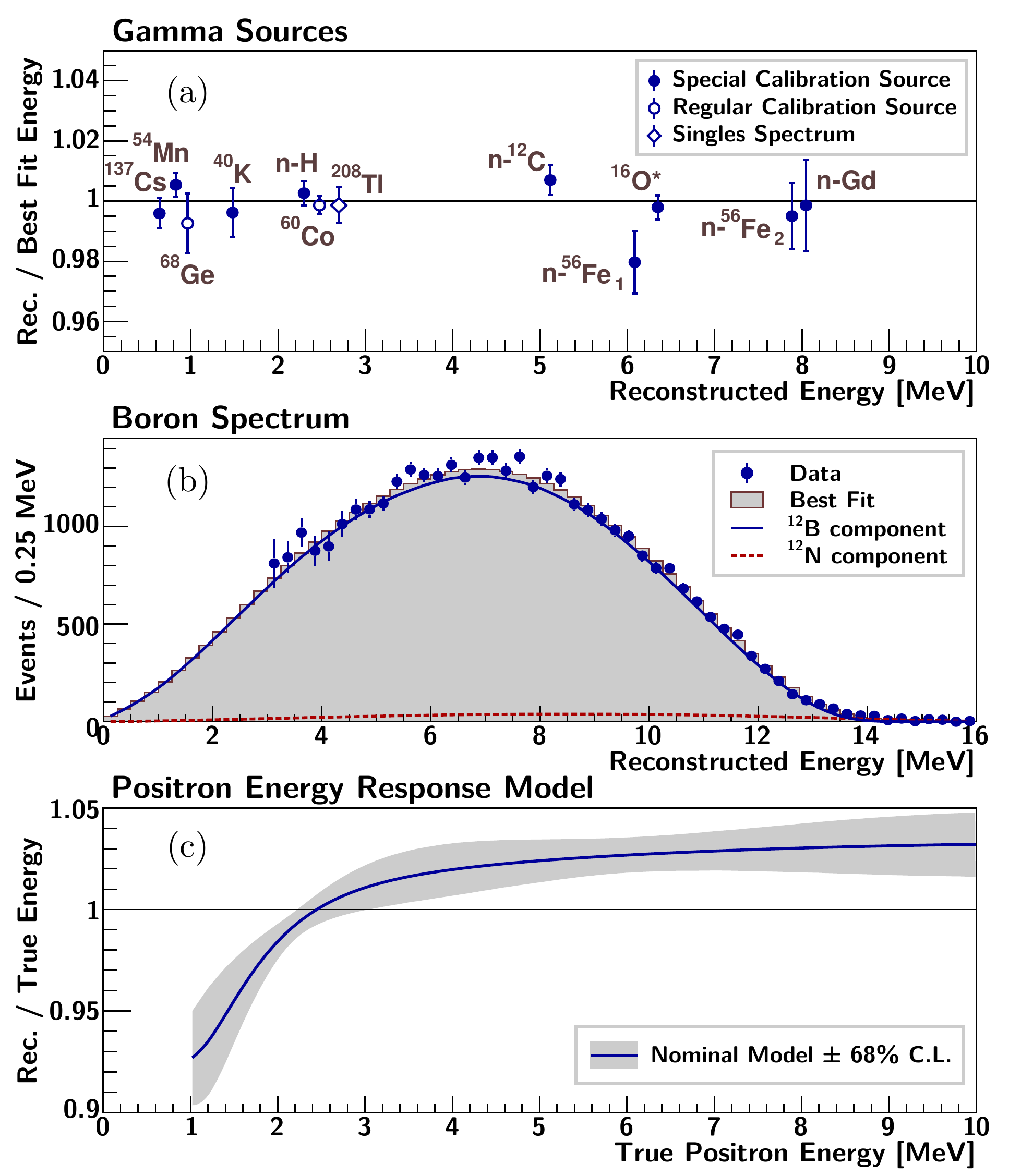} 
 \end{tabular}
 \caption{Left panel: three zones Daya Bay antineutrino detector. Right panel:(a) Ratio of the reconstructed to 
best-fit energies of $\gamma$ lines from calibration sources and singles spectra. The total uncertainty on each ratio 
is shown as the error bars. (b) Reconstructed energy spectrum (points) compared to the sum (shaded area) of the $^{12}$B 
(solid line) and $^{12}$N (dashed line) components of the best-fit energy response model. The error bars represent the 
statistical uncertainties. (c) AD energy response model for positrons. }
 \label{fig:AD}
\end{figure}

The Daya Bay experiment is described in details  in ~\cite{DayaBay:2012aa,Guo:2007ug}. Here we will only briefly recall 
the main points. There are three experimental halls (EHs) -- two ``near'' and one ``far'' which contain functionally 
identical, antineutrino detectors (ADs) surrounded by a pool of ultra-pure water segmented into two regions, 
the inner water shield and outer water shield, which are instrumented with photomultiplier tubes (PMTs). 
The Daya Bay experiment uses three-zones antineutrino detectors (AD) schematically shown in the left 
panel of Fig.~\ref{fig:AD}. 

The inner zone is filled by 20 tons of gadolinium (Gd) doped liquid scintillator (LS) contained in acrylic vessel. The 
middle zone is filled by 20 tons of LS without gadolinium contained in acrylic vessel. The outer zone is filled by 40 
tons of mineral oil. Both inner and middle zones are used to detect $\bar{\nu}_e$ via inverse beta decay (IBD) reaction 
$\bar{\nu}_e + p\to e^+ + n$. The IBD identification exploites the time structure of the IBD event -- a prompt signal 
due to $e^+$ energy loss and subsequent annihilation with $e^-$ is followed by a recoil neutron capture. The neutron 
can be captured by Gd nucleus or by proton.  We call the corresponding analyses as ''Gd analysis`` and ''nH analysis`` 
respectively. Apparently, the inner zone of AD is the only target of $\bar{\nu}_e$ for the nGd analysis while both inner 
and middle zones serve as the targets for the nH analysis. The outer zone is used to suppress the background and 
external radioactivity from PMT and stainless steel structures. Also it suppresses the scintillation in the outer 
region. The inside volume is viewed by 192 8-inch Hamamatsu PMTs. On average 1 MeV of released energy inside of LS 
corresponds to about 163 photoelectrons detected by PMTs. The energy resolution is 
estimated as $(7.5/\sqrt{E_\text{vis}/\text{MeV}} + 0.9)\%$. The ADs are a subject of systematic calibrations compaigns 
regularly checking the energy response of ADs.

Interpretation of the observed prompt energy spectra requires mapping of the detector response to $e^+$, $e^-$ and 
$\gamma$ with the true released visible energy ($E_{\rm true}$) to the reconstructed energy ($E_{\rm rec}$). $E_{\rm 
rec}$ is determined by scaling the measured total charge with a position-dependent 
correction~\cite{DayaBay:2012aa,An:2013uza}. Right panel of Fig.~\ref{fig:AD} compares the best-fit energy model with 
the single-gamma, multi-gamma and continuous $^{12}$B   data used to determine the model parameters. As additional 
validation, the energy model prediction for the continuous $\beta+\gamma$ spectra from $^{212}$Bi, $^{214}$Bi and 
$^{208}$Tl decays was compared with the data and found to be consistent.

\section{\label{sec:results}Results}
\subsection{Neutrino oscillation analyses}

\begin{figure}[htb]
\begin{tabular}{cc}
\centering\includegraphics[width=0.5\linewidth]{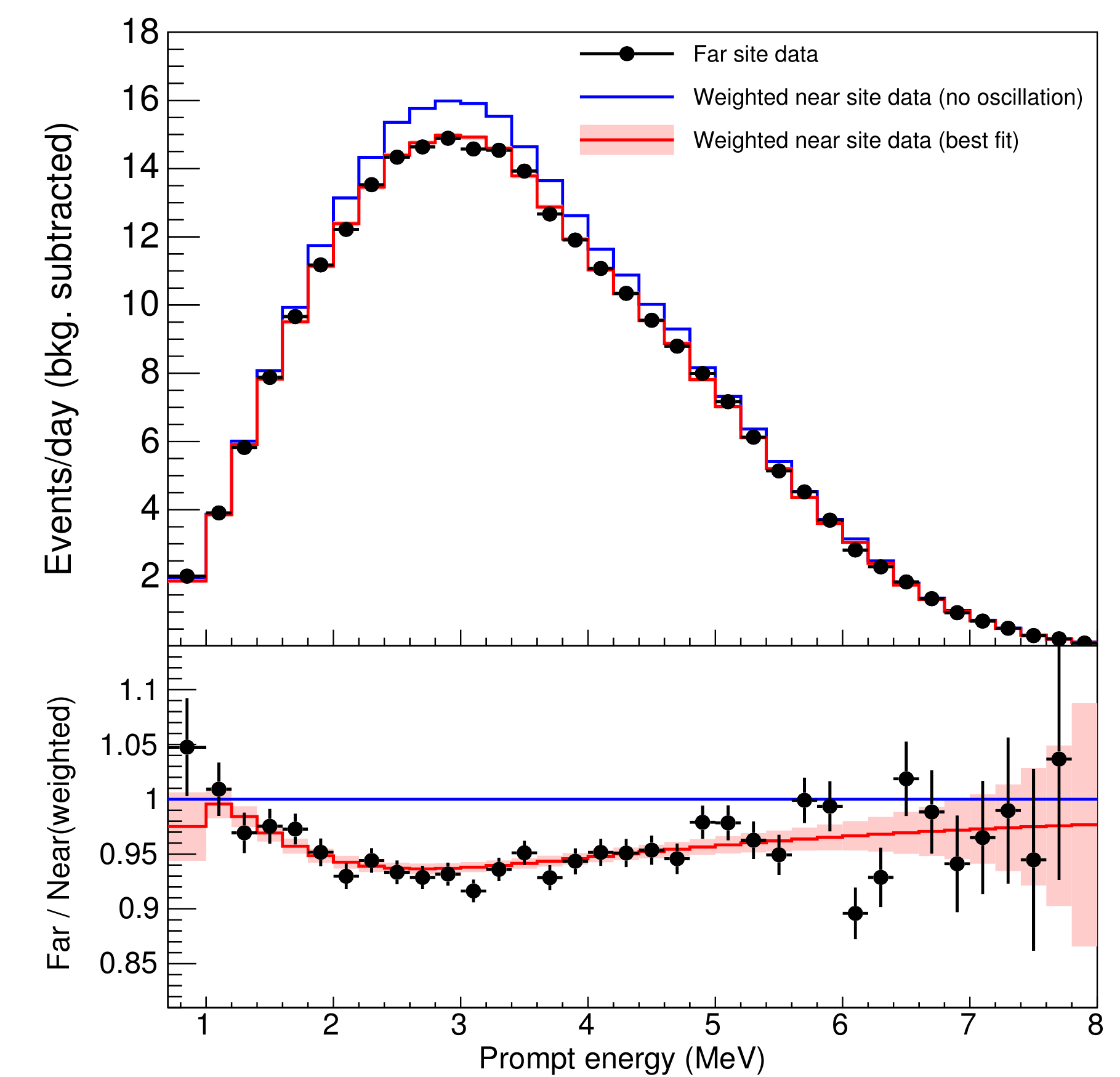}&
\centering\includegraphics[width=0.45\linewidth]{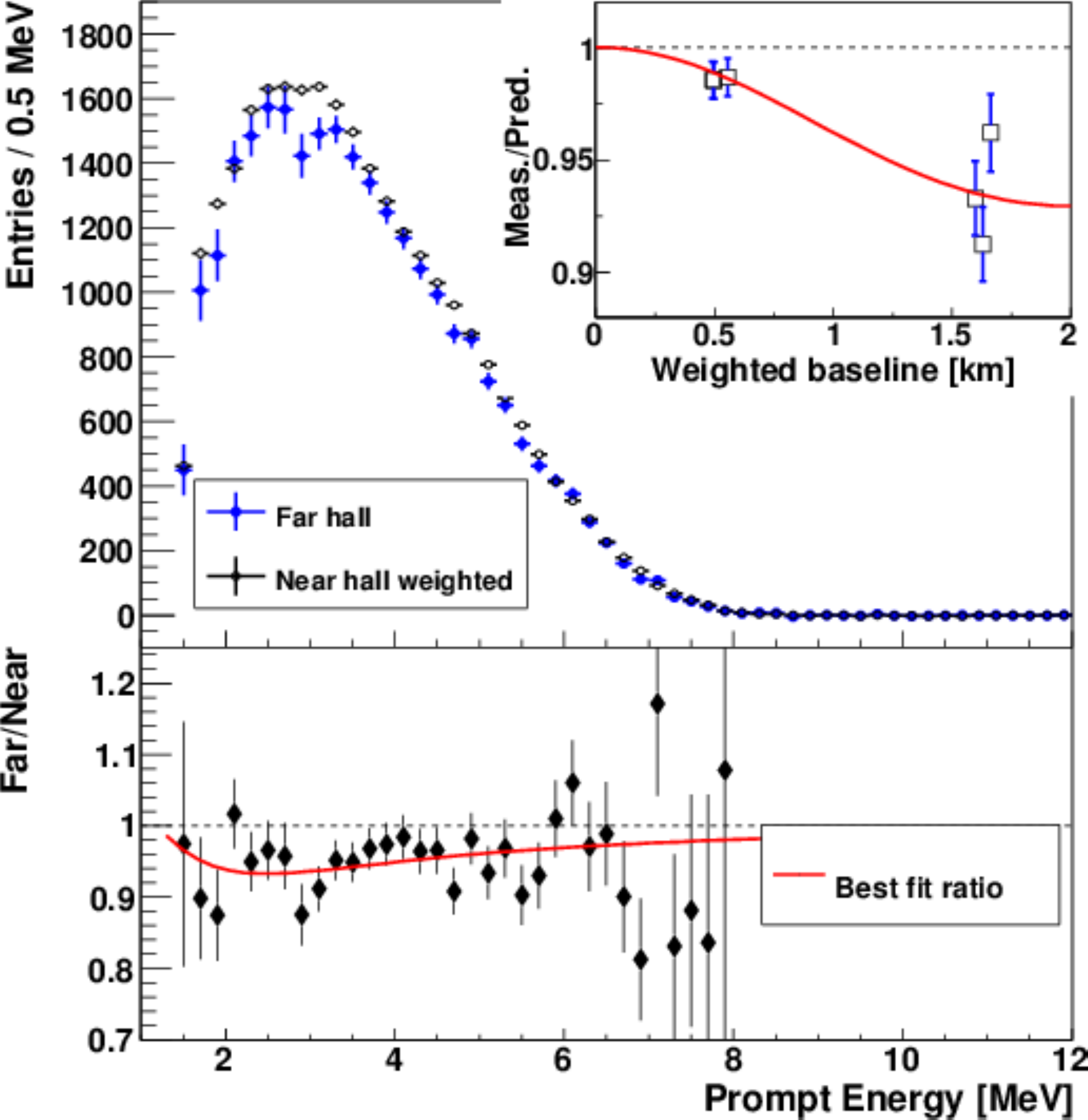}
\end{tabular}
\caption{Left panel: nGd analysis. The upper panel shows the background substracted prompt positron spectra (black 
points) measured in the far experimental hall. The blue line shows the expectations  based on the near site data 
assuming no oscillation. The red line represents the expectations based on the near site data assuming best fit 
oscillation. The band corresponds to the uncertainty in the prediction. In the lower panel the black points represent 
the ratio of the background-subtracted data divided by the predicted no-oscillation spectra. The error bars represent 
the statistical uncertainty only. The red curve in each lower panel represents the ratio of the best-fit to 
no-oscillations spectra. The change in slope of the red curve in the lowest energy bin is due to the effect of energy 
loss in the acrylic. Right panel: nH analysis. The detected energy spectrum of the prompt events of the far hall ADs 
(blue) and near hall ADs (open circle) weighted according to baseline. The far-to-near ratio (solid dot) with best fit 
$\theta_{13}$ value is shown in the lower plot. In the inset is the ratio of the measured to the predicted rates in each 
AD vs baseline, in which the AD4 (AD6) baseline was shifted relative to that of AD 5 by 30 (-30) m for visual clarity.}
\label{fig:oscillation_new}
\end{figure}
Two more oscillations analyses have been carried out since publications~\cite{An:2012eh,An:2013uza}. First is nGd 
analysis~\cite{An:2013zwz} and the second is nH analysis with largely independent systematics and event 
selection~\cite{An:2014ehw}. Let us begin with a discussion of the results of the first analysis. The rate uncertainty 
of the background is slightly reduced compared to the previous analyses~\cite{An:2012eh,An:2013uza} due to the increased 
statistics. The analysis includes energy shape information by applying the energy nonlinearity correction shown in 
the right panel of Fig.~\ref{fig:AD} to the positron spectrum and measuring the energy shape distribution of the five 
background sources. The spectral uncertainties of the five backgrounds are included as uncorrelated among energy bins in 
the $\chi^2$ fit of the oscillation parameters, to allow all possible spectral models consistent with the data. The 
combined rate and spectral analysis yields $\sin^22\theta_{13} = 0.084\pm 0.008$ and 
$|\Delta m^2_{ee}| = (2.44^{+0.10}_{-0.11})\times 10^{-3} {\rm eV}^2$ with $\chi^2/{\rm NDF} = 134.7/146$. The 
corresponding measured prompt energy spectrum is compared to the expectations assuming no oscillations and oscillations 
with best fit parameters as measured by Daya Bay is shown in the left panel of Fig.~\ref{fig:oscillation_new}.

Let us now discuss the nH analysis. 217 days of data taking corresponding to the Daya Bay time period when only  6 ADs 
were functioning have been used in this analysis. Since $n+p\to {}^2\text{D}+\gamma$ 
reaction releases smaller energy ($2.2$ MeV) than Gd excitations (about $8$ MeV) there are more accidentals due to the 
lower delayed energy threshold. This is one example of generally somewhat different systematics relative to nGd 
analysis. 

As a result of nH analysis the far detectors also observe a deficit in the event rate compared to the expectations 
based on near detectors measurements. Within the three-neutrino oscillation framework it allows us to measure $\sin^2
 2\theta_{13} = 0.083 \pm  0.018$ in good agreement with nGd analysis. While nH spectral analysis is in progress one 
could observe that the spectral distortion is consistent with oscillations as can be seen from the right panel of 
Fig.~\ref{fig:oscillation_new}.

\subsection{Absolute reactor neutrino flux measurement}
\begin{figure}[htb]
\begin{center}
\includegraphics[width=0.8\linewidth]{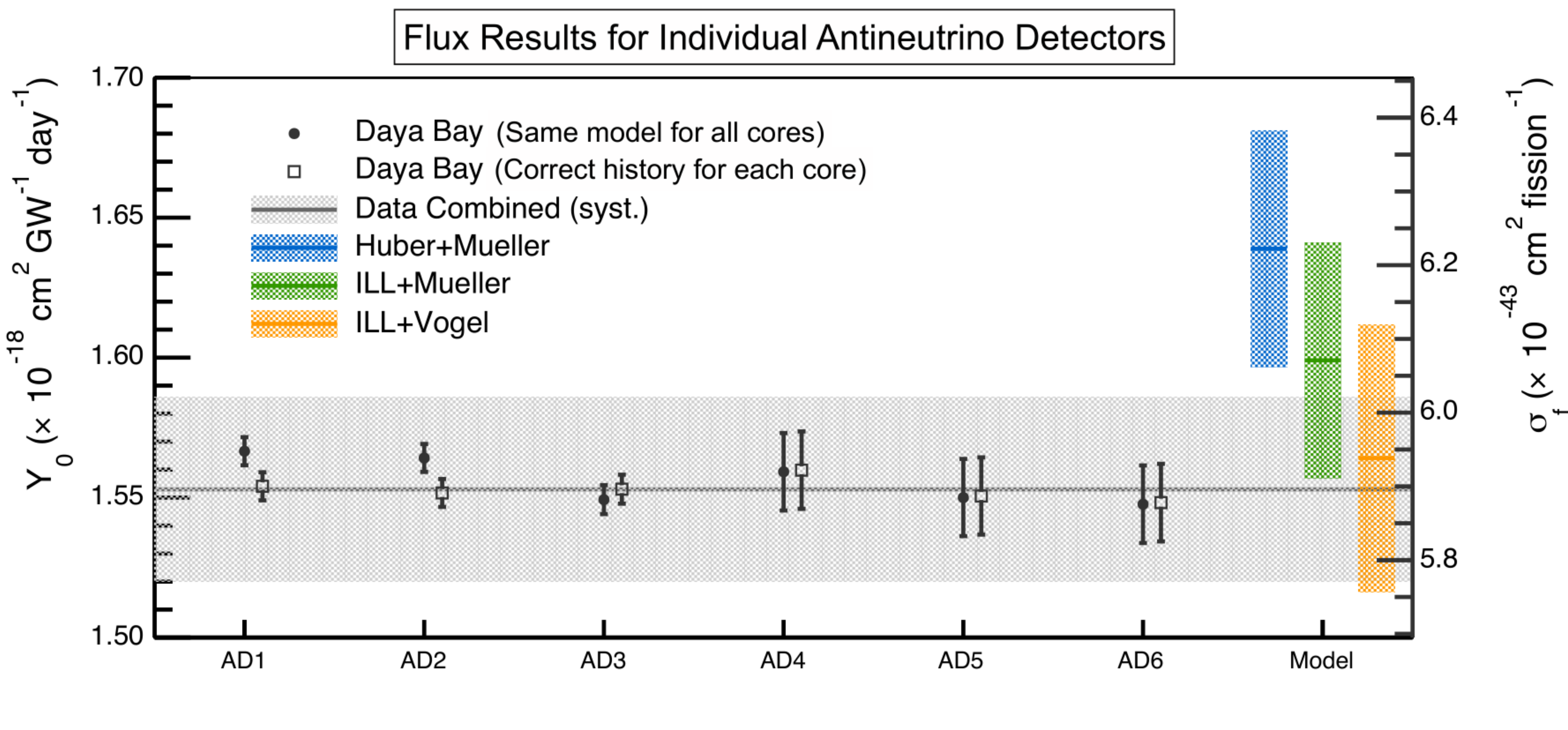}  
\caption{The measured reactor $\bar{\nu}_e$ event rate at each AD after correcting for the $\bar{\nu}_e$ survival 
probability, re-expressed as $Y_0$ (cm$^2$ GW$^{-1}$ day$^{-1}$) and $\sigma_f$ (cm$^2$ fission$^{-1}$). The solid and 
open circles show the data without and with correction for the difference in the effective fission fractions observed 
by each AD. The uncertainty of the measurement is shown as the gray band. Three theoretical model predictions are shown 
as a reference.}
\label{fig:absrate_ads}
\end{center}
\end{figure}

The large reactor $\bar{\nu}_e$ sample collected at Daya Bay allows for a precise measurement of the absolute 
reactor antineutrino flux. The analysis uses the complete 217-day data set of the 6-AD period. A total of 300k (40k) 
candidates are detected at the near (far) halls. Fig. \ref{fig:absrate_ads} shows the measured reactor $\bar{\nu}_e$ 
event rate at each AD after correcting for the $\bar{\nu}_e$ survival probability, re-expressed as $Y_0$ (cm$^2$ 
GW$^{-1}$ day$^{-1}$) and $\sigma_f$ (cm$^2$ fission$^{-1}$). The measurement among ADs is consistent within statistical 
fluctuations after correcting for the difference in the effective fission fractions. The uncertainty (2.3\%) of the 
measurement is dominated by the uncertainty in detection efficiency (2.1\%), which is correlated among all ADs. The 
measurement yields an average $Y_0 = 1.553 \times 10^{-18} \text{cm}^2 \text{GW}^{-1} \text{day}^{-1}$ and $\sigma_f = 
5.934 \times 10^{-43} \text{cm}^2 \text{fission}^{-1}$, with the average fission fractions 
${}^{235}\text{U}:{}^{238}\text{U}:
{}^{239}\text{Pu}: {}^{241}\text{Pu} = 0.586 : 0.076 :0.288 : 0.050$.
Three theoretical model predictions are shown in Fig.~\ref{fig:absrate_ads} as a reference. The 
Huber~\cite{Huber:2011wv} and ILL~\cite{Schreckenbach:1985ep,Hahn:1989zr} models predict the $\bar{\nu}_e$ spectra for 
${}^{235}\text{U}$, ${}^{239}\text{Pu}$ and ${}^{241}\text{Pu}$, while the Mueller~\cite{Mueller:2011nm} and 
Vogel~\cite{Vogel:1980bk} models predict for ${}^{238}\text{U}$. The uncertainty in the model predictions is 
estimated by authors to be $\simeq 2.7$\%. This estimate might be somewhat optimistic as follows 
from~\cite{Hayes:2013wra} which suggests the corresponding uncertainty to be not less than $4\%$. The ratio (R) of the 
Daya Bay measurement to the Huber+Muller model prediction is $R = 0.947\pm 0.022$, while $R = 0.992\pm 0.023$ when 
compared to the ILL+Vogel model prediction.

\begin{figure}[htb]
\begin{center}
\includegraphics[width=0.8\linewidth]{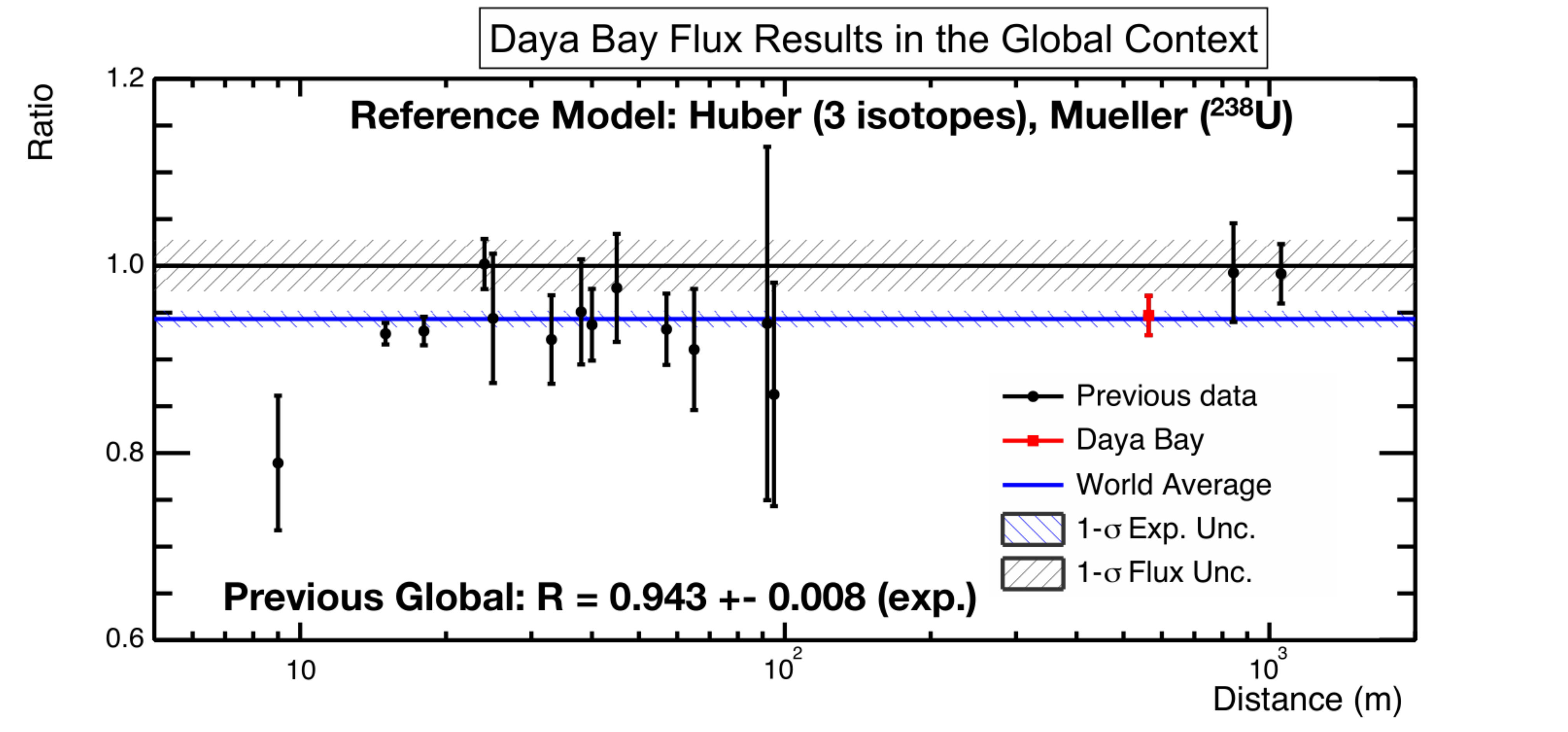} 
\end{center}
\caption{The reactor $\bar{\nu}_e$ interaction rate of the 21 previous 
short-baseline experiments~\cite{Mention:2011rk,Zhang:2013ela} as a function of the distance from the reactor, 
normalized to the Huber+Mueller model prediction~\cite{Huber:2011wv, Mueller:2011nm}. Experiments at the same baseline 
are combined together for clarity. The Daya Bay experiment is placed at the effective baseline of 573 m. The rate is 
corrected by the $\bar{\nu}_e$ survival probability at the distance of each experiment, assuming standard three-neutrino 
oscillation. The horizontal bar (blue) represents the global average and its $1\sigma$ uncertainty. The $2.7$\% reactor 
flux uncertainty is shown as a band around unity.}
\label{fig:absrate_ratio}
\end{figure}
The Daya Bay result is compared to the 21 past reactor neutrino flux measurements as shown in  
Fig.~\ref{fig:absrate_ratio} according to Refs.~\cite{Mention:2011rk,Zhang:2013ela}. As the  common reference model for 
all experiments in Fig.~\ref{fig:absrate_ratio} the Huber+Mueller model is used assuming the neutron lifetime 
value to be 880.1 s ~\cite{Beringer:1900zz}. The $\bar{\nu}_e$ survival probability is calculated with 
$\sin^22\theta_{13} = 0.089\pm 0.009$ determined from the rate-only analysis~\cite{An:2013zwz}. The global average of 
the 21 past measurements with respect to the Huber+Mueller model prediction is determined to be  $R =0.943\pm 0.008$ 
(experimental uncertainty), which is consistent with $R = 0.947\pm 0.022$ from the Daya Bay measurement.

\subsection{Energy spectrum measurement}
\begin{figure}[htb]
 \centering\includegraphics[width=0.5\textwidth]{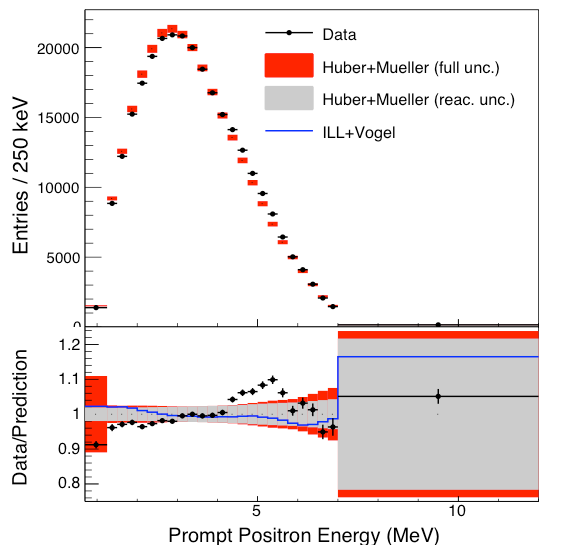}
 \caption{Upper panel: the measured prompt energy spectrum by Daya Bay experiment compared to the expectations based on 
Huber+Muller model prediction~\cite{Huber:2011wv,Mueller:2011nm}. Bottom panel: data/prediction ratio. The 
shadowed area respresents the theory model estimation of the uncertainty.}
\label{fig:spectrum_measurement}
\end{figure}

A preliminary comparison of the measured prompt energy spectrum to the expectations based on Huber+Muller model 
prediction~\cite{Huber:2011wv,Mueller:2011nm} is displayed in Fig.~\ref{fig:spectrum_measurement}. 

% \begin{wrapfigure}{l}{0.5\textwidth}
%  \centering\includegraphics[width=0.5\textwidth]{spectrum_measurement}
%  \caption{Upper panel: the measured prompt energy spectrum by Daya Bay experiment compared to the expectations based on 
% Huber+Muller model prediction~\cite{Huber:2011wv,Mueller:2011nm}. Bottom panel: data/prediction ratio. The 
% shadowed area respresents the theory model estimation of the uncertainty.}
% \label{fig:spectrum_measurement}
% \end{wrapfigure}
One can observe a significant mismatch of the spectra in the energy region $4-6$ MeV where the local significance of 
the discrepancy 
reaches the level of $4\sigma$. This excess is observed also by RENO~\cite{Seon-HeeSeofortheRENO:2014jza} and Double 
Chooz~\cite{Crespo-Anadon:2014dea} reactor experiments. The excess is unlikely to be caused by unaccounted for detector 
effects or additional background. It matches all characteristics of IBD events. It correlates to the reactor power and 
apart of that is time independent. First-principle calculations of fission and $\beta$ decay processes predict similar 
excess~\cite{Dwyer:2014eka} where the authors conclude ``The presence of this bump in both the calculated electron 
and antineutrino spectra suggests that the discrepancy may not be due to systematics of the $\beta^-$ conversion 
method, but instead may be an artifact of the original $\beta^-$ measurements''. Today the origin of this descrepancy 
is an open question.

\subsection{Light sterile neutrino search}
\begin{figure}[htb]
\begin{tabular}{cc}
 \includegraphics[width=0.55\textwidth]{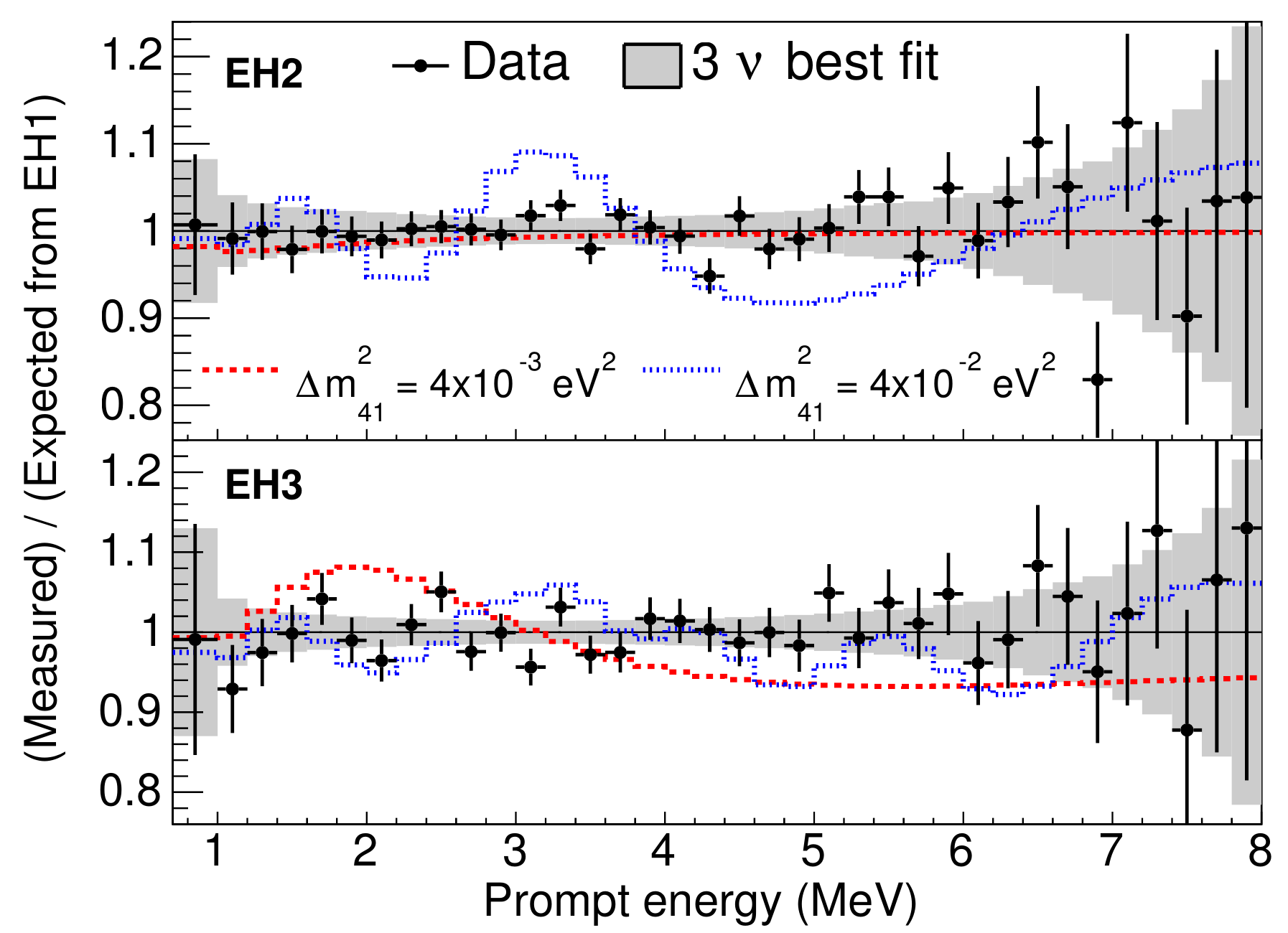}&
 \includegraphics[width=0.4\textwidth]{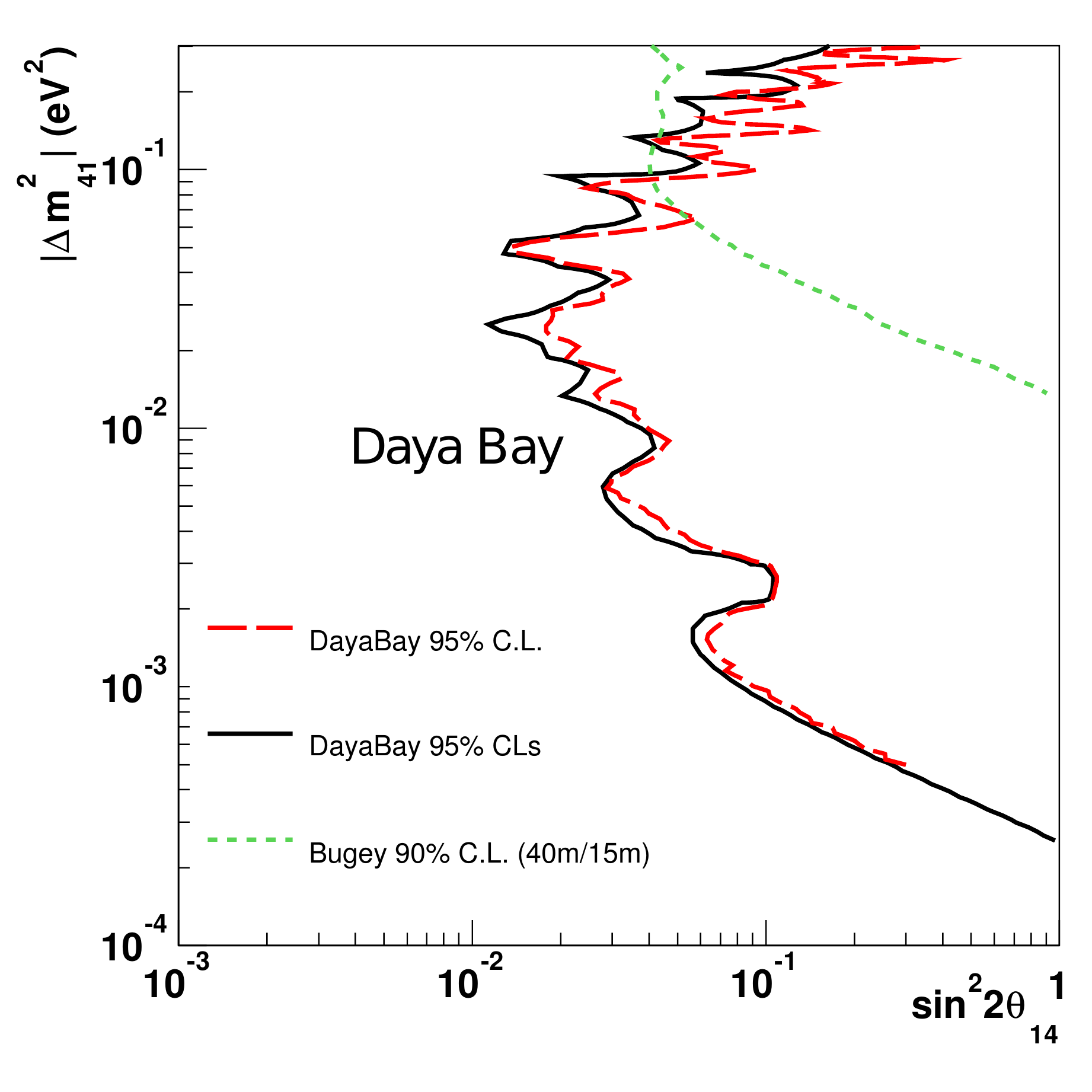}
\end{tabular}
\caption{Left panel: Prompt energy spectra observed at EH2 (top) and EH3 (bottom), divided by the prediction from the 
EH1 spectrum with the three-neutrino best fit oscillation parameters from the previous Daya Bay analysis 
~\cite{An:2013zwz}. The gray band represents the uncertainty of three-neutrino oscillation prediction, which includes 
the statistical uncertainty of the EH1 data and all the systematic uncertainties. Predictions with $\sin^22\theta_{14} 
= 0.1$ and two representative $|\Delta m^2_{41}|$ values are also shown as the dotted and dashed curves. Right panel: 
The exclusion contours for the neutrino oscillation parameters $\sin^{2}2\theta_{14}$ and $|{\Delta}m^{2}_{41}|$. 
Normal mass hierarchy is assumed for both $\Delta m_{31}^{2}$ and $\Delta m_{41}^{2}$.  The red long-dashed curve 
represents the 95\% C.L. exclusion contour with Feldman-Cousins method~\cite{Feldman:1997qc}. The black solid curve 
represents the 95\% ${\rm CL_{s}}$ exclusion  contour~\cite{Read:2002hq}. The parameter-space to the right side of the 
contours are excluded. For comparison, Bugey's~\cite{Declais:1994su} 90\% C.L. limit on $\overline{\nu}_e$ 
disappearance is also shown as the green dashed curve. 
}
\label{fig:sterile_spec}
\end{figure}

The Daya Bay experiment performed a search for a possible sterile neutrino. What is the sterile neutrino? It is a 
quantum state defined as a coherent (``flavor'')  mixture of massive states $\nu_1,\nu_2,\nu_3,\nu_4, \text{etc}$ which 
does not interact with $W^\pm,Z$. However each of massive $\nu_i$ does interact with gauge bosons. The $4\times 4$ (in 
a minimal extension of the Standard Model) unitary mixing matrix is organized in such a way that four massive neutrinos 
contribute as just three states to the widths of $W^\pm,Z$. However it does not mean that fourth (or more) massive 
neutrino remains invisible. If an initially produced flavour state ($\nu_e,\nu_\mu,\nu_\tau$) evolves with time it might 
appear as sterile state thus making additional deficit of the detected events. In Daya Bay the sterile neutrino could 
cause additional spectral distortion betweens the ADs thanks to multiple baselines (350 m, 500 m  and 1600 m) as shown 
in the left panel of Fig.~\ref{fig:sterile_spec}.
The analysis uses the complete 217-day data set of the 6-AD period. The relative spectral distortion due to the 
disappearance of $\bar{\nu}_e$ is found to be consistent with that of the three-flavor oscillation model. The exclusion 
contours for $\sin^22\theta_{14}$ and $|\Delta m^2_{41}|$ displayed in the right panel of 
Fig.~\ref{fig:sterile_spec} are determined using both the Feldman-Cousins method~\cite{Feldman:1997qc} and the CLs 
method~\cite{Qian:2014nha}. The derived limits cover the $10^{-3} \text{ eV}^2 < |\Delta m^2_{41}| < 0.1 \text{ eV}^2$ 
region, which was previously largely unexplored. Details of the sterile neutrino analysis can be found in 
~\cite{An:2014bik}.
        
\section{Summary}
\label{sec:summary}
The Daya Bay experiment uses the relative measurement of the $\bar{\nu}_e$ rate and spectrum between near and far 
detectors to precisely measure the oscillation parameters $\sin^22\theta_{13}$ and $|\Delta m^2_\text{ee}|$. The Daya 
Bay experiments takes the data in the final 8-AD configuration since summer 2012 when two new ADs were installed. With 
621 days of data, Daya Bay has measured $\sin^22\theta_{13}=0.084 \pm 0.005$ and $|\Delta 
m^2_\text{ee}|=2.44^{+0.10}_{-0.11}\times 10^{-3}\text{ eV}^2$. This is the most precise measurement of 
$\sin^22\theta_{13}$ to date. The precision measurement of $\theta_{13}$ opens the door for future experiments to study 
neutrino mass hierarchy and leptonic CP violation. The $|\Delta m^2_\text{ee}|$ measurement is in agreement with 
$|\Delta m^2_{\mu\mu}|$ measurements by the muon neutrino disappearance experiments. The precisions of both $|\Delta 
m^2_\text{ee}|$ and $|\Delta m^2_{\mu\mu}|$ measurements are comparable today. By the end of 2017 Daya Bay expects to 
measure both $\sin^22\theta_{13}$ and $|\Delta m^2_\text{ee}|$ with precisions better than 3\%. 

Several other analyses have been also performed. $\theta_{13}$ angle has been measured in nH analysis yielding  
$\sin^22\theta_{13}=0.083 \pm 0.018$. The absolute reactor antineutrino flux measurement has yielded results consistent 
with previous short-baseline reactor neutrino experiments thus confirming the ``Reactor Antineutrino Anomaly'' 
first introduced in Ref.~\cite{Mention:2011rk}. However it is still an open question if the anomaly is due to sterile 
neutrinos or due to  uncertainties in the model calculations of reactor antineutrino fluxes. Therefore, an analysis 
based mostly on the energy shape information and exploiting multiple baselines of Daya Bay experiment has been 
performed searching for a possible signal of sterile neutrino in the observed energy spectra. Such a signal has not 
been observed which allows us to set stringent limits in the $10^{-3} \text{ eV}^2 < |\Delta m^2_{41}| < 0.1 \text{ 
eV}^2$ region.  Finally, preliminary results on energy spectrum of reactor antineutrino show generally a good 
agreement with expectations~\cite{Huber:2011wv,Mueller:2011nm} except the energy interval $4-6$ MeV with where a 
significant mismatch has been observed.
\section*{ACKNOWLEDGMENTS}
Daya Bay is supported in part by the Ministry of Science and Technology of China, the U.S. Department of Energy,
the Chinese Academy of Sciences, the National Natural Science Foundation of China, the Guangdong provincial government,
the Shenzhen municipal government, the China General Nuclear Power Group, Key Laboratory of Particle
and Radiation Imaging (Tsinghua University), the Ministry of Education, Key Laboratory of Particle Physics and
Particle Irradiation (Shandong University), the Ministry of Education, Shanghai Laboratory for Particle Physics and
Cosmology, the Research Grants Council of the Hong Kong Special Administrative Region of China, the University
Development Fund of The University of Hong Kong, the MOE program for Research of Excellence at National Taiwan
University, National Chiao-Tung University, and NSC fund support from Taiwan, the U.S. National Science Foundation,
the Alfred P. Sloan Foundation, the Ministry of Education, Youth, and Sports of the Czech Republic, the Joint
Institute of Nuclear Research in Dubna, Russia, the CNFC-RFBR joint research program, the National Commission of
Scientific and Technological Research of Chile, and the Tsinghua University Initiative Scientific Research Program.
We acknowledge Yellow River Engineering Consulting Co., Ltd., and China Railway 15th Bureau Group Co., Ltd.,
for building the underground laboratory. We are grateful for the ongoing cooperation from the China General Nuclear
Power Group and China Light and Power Company.

I would like also to warmly thank all my collegues from the Daya Bay Collaboration and especially Maxim Gonchar, S\"oren 
Jetter, Jiajie Ling, Logan Lebanowski, Pedro Ochoa, Xin Qian, Wei Wang, Zhe Wang, Chao Zhang, Weili Zhong for useful 
discussions and providing me with supplemental material for this manuscript.

\end{document}